\documentclass[conference]{IEEEtran}
\usepackage{cite}
\usepackage{amsmath,amssymb,amsfonts}
\usepackage{pifont}
\usepackage{algorithmic}
\usepackage{graphicx}
\usepackage{multirow}
\usepackage{xspace}
\usepackage{textcomp}
\usepackage[table,xcdraw]{xcolor}
\usepackage{balance}
\usepackage{paralist}
\usepackage[most]{tcolorbox}
\usepackage{diagbox}
\usepackage{hhline}
\usepackage[noblocks]{authblk}

\def\BibTeX{{\rm B\kern-.05em{\sc i\kern-.025em b}\kern-.08em
    T\kern-.1667em\lower.7ex\hbox{E}\kern-.125emX}}
    
\usepackage{xcolor}
 
\usepackage{hyperref}
\hypersetup{
	breaklinks=true,
	colorlinks=true,
	linkcolor=blue,
	citecolor=blue,
	urlcolor=blue,
}
\usepackage{breakurl}
\usepackage[flushleft]{threeparttable}
\usepackage{todonotes}

\newcommand{\subhead}[1]{\vspace {1pt}\noindent{\textbf{#1.}}}

\def\showcomments{1}
\ifnum\showcomments=1
    \newcommand{\fixme}[1]{{\textcolor{red}{[FIXME: #1]}}}
    \newcommand{\checkme}[1]{{\textcolor{orange}{[CHECKME: #1]}}}
    \newcommand{\ssnote}[1]{{\textcolor{blue}{[SS: #1]}}}
    \newcounter{mynote}[section]
    
    \newcommand{\thenote}{\thesection.\arabic{mynote}}
    \newcommand{\viau}[1]{
         \textcolor{teal}{Viau: #1}
}
\else
    \newcommand{\fixme}[1]{}
    \newcommand{\checkme}[1]{}
    \newcommand{\thenote}[1]{}
    \newcommand{\ssnote}[1]{}
    \newcommand{\viau}[1]{}
\fi

\begin{document}
\title{SoK: Hardware Security Support for Trustworthy Execution
}
\author{Lianying Zhao}
\affil{Carleton University\textsuperscript{$\ddagger$}, Ottawa, ON, Canada}
\author{He Shuang}
\author{Shengjie Xu}
\author{Wei Huang}
\author{Rongzhen Cui}
\author{Pushkar Bettadpur}
\author{David Lie}
\affil{University of Toronto, Toronto, ON, Canada}

\maketitle

\begin{abstract}
In recent years, there have emerged many new hardware mechanisms for improving the security of our computer systems.  Hardware offers many advantages over pure software approaches: immutability of mechanisms to software attacks, better execution and power efficiency and a smaller interface allowing it to better maintain secrets.  This has given birth to a plethora of hardware mechanisms providing trusted execution environments (TEEs), support for integrity checking and memory safety and widespread uses of hardware roots of trust.  

In this paper, we systematize these approaches through the lens of \textit{abstraction}.  Abstraction is key to computing systems, and the interface between hardware and software contains many abstractions.  We find that these abstractions, when poorly designed, can both obscure information that is needed for security enforcement, as well as reveal information that needs to be kept secret, leading to vulnerabilities.  We summarize such vulnerabilities and discuss several research trends of this area.

\end{abstract}


\section{Introduction} \label{sec:intro}

\let\oldthefootnote\thefootnote
\renewcommand{\thefootnote}{$\ddagger$}
\footnotetext{The work was started when the first author was at the University of Toronto.}
\renewcommand{\thefootnote}{\oldthefootnote}

The trustworthiness of a computer system 
entails that the system can correctly fulfill tasks as intended by the user.
This can include user tasks (e.g., applications) and system tasks (e.g., a system integrity monitor). Enabling techniques for trustworthy execution can be implemented in software, hardware or both. For instance, there exist proposals for both software-based~\cite{cfi-sw} and hardware-based~\cite{cfi-hw} solutions for Control-Flow Integrity (CFI); and likewise, isolation for code can be achieved with either software (e.g., containerization and sandboxing) or hardware components~\cite{iso}.

Nonetheless, there exists a common belief in hardware's advantages over software
in consideration of the following aspects:
\begin{inparaenum}
\item Relative immutability. Tampering with hardware is non-trivial and requires physical access.
This makes hardware support a necessity in mitigation techniques under a strong adversarial model, e.g., rootkit-level threats~\cite{ki-mon} and hypervisor threats~\cite{vmbr}.
\item Efficiency. 
Contrary to software that introduces a layer of architectural abstraction, direct hardware implementation can save redundant processing cycles (e.g., instruction decode).
This allows for better efficiency for
certain iterative/repetitive operations (e.g., consider how DSPs outperform regular CPUs for signal processing). 
It can further contribute to lowering power consumption, which is 
critical for resource-constrained devices. 
\end{inparaenum}

Furthermore, hardware is the \emph{Root of Trust} (RoT)~\cite{rot}, as it bridges the physical world (where human users reside) and the digital world (where tasks run as software). To securely perform a task or store a secret, the user trusts at least part of the computer hardware.

Dedicated hardware security support has seen its proliferation since the early days of computers. It can take a straightforward form as discrete components to assist the CPU, ranging from
the industrial-grade tamper-responding IBM Cryptocards (e.g., 4758~\cite{ibm4758}), Apple's proprietary secure enclave processor (SEP~\cite{sep}) for consumer electronics, to the more open Trusted Platform Module (TPM), smart cards and other security tokens.
Meanwhile, architectural enhancements are being introduced to the CPU, as well as its chipset, such as Trusted Execution Environments (TEEs, see Section~\ref{sec:anchor}), the NX-bit, Intel MPX, MPK, etc.

Driven by demand over the few decades, hardware is being ``patched'' cumulatively, as reflected in the increase of the x86 ISA extensions (according to Baumann~\cite{newsoftware}, hardware is becoming the new software). The added complexity and interaction between hardware features has led to numerous problems, which has motivated us to reflect on the role of hardware in security, and what key features make a good hardware security mechanism.

Most hardware security features achieve their purpose by hiding information or access from an entity, and/or exposing a new/reduced interface to it. For instance, memory protection between processes is implemented by virtual memory, which prevents processes from accessing each other's memory by presenting each process with its own virtual address space.  The design of virtual memory can be considered an instance of ``abstraction''~\cite{abstraction,abstraction2}, which 1) removes or attenuates information (by hiding the true implementation of memory) and 2) creates new semantics (such as page tables), which are also called ``abstractions''. Hence there is a relationship between security and abstraction, which is particularly important considering the heavy use of abstractions in computer hardware to separate its implementation from that of the software it runs. 

We propose to use abstraction (Section~\ref{sec:abs}) as a lens through which the workings of hardware security support and various related issues are examined. 
Specifically, we find that:
\begin{inparaenum}[1)]\
\item beyond the commonly perceived immutability and efficiency, the properties of hardware security support can be explained as mechanisms that hide/expose something; 
\item certain low-level attacks can be attributed to insufficient (Section~\ref{sec:under}) or excess (Section~\ref{sec:over})  abstraction.; and 
\item many proposals seek to leverage abstractions to increase security (Section~\ref{sec:ee}) or propose new abstractions to increase the flow of information between hardware and software for better security (Section~\ref{sec:misc}).
\end{inparaenum}

While traditionally Trusted Computing~\cite{trustedcomputing} emphasizes merely the protection for initial integrity and isolation during execution, we expand the scope to be more holistic and
to ensure the eventual correct \emph{execution outcome}.
This includes run-time dynamic aspects such as 
control/data flow integrity of the running program and secure input/output.  We also include firmware in the processor, system chipset and peripheral devices, which while they do not execute trusted application software, are still nonetheless part of the trusted computing base of much of that software.
We examine how hardware-enforced abstraction affects execution within this entire scope.

It is not our intention to focus on improper implementation that fails to properly implement a specification. However, a hardware-related fault may not clearly be attributable to either a design flaw or an implementation error. We will not try to draw a boundary (which might be difficult to do in any case).

\vspace{5pt}
\subhead{Contributions}
\begin{enumerate}
\item We develop a general model for hardware security support based on the notion of abstraction, and examine various Trusted Execution Environments (TEEs) along several key properties of the abstractions the TEEs provide.
\item We associate the extent of abstraction (excessive/insufficient) with the root cause of a series of vulnerabilities, represented by side channels on one hand and insufficient information flow on the other.
Namely, when the abstraction specification does not hide all that is supposed to be hidden, leaked information leads to attacks. On the other hand, when such leakage is inevitable, too much hiding across components also hinders effective reference monitoring, which in turn leads to compromises.
\item We systematize the application of state-of-the-art hardware security features and the proposal of new hardware enhancements in the literature, based on their properties from the perspective of abstraction.
\end{enumerate}

\section{Model of Trustworthy Execution} \label{sec:model}

In this section, we explain our model of trustworthy execution, using  abstraction as a way to  analyze and understand the various hardware support towards trustworthy execution  that has been implemented over the years.  We begin by defining abstraction in this context and show how it maps to the operation of hardware.  We then discuss its role in improving hardware support for trustworthy execution.

\subsection{Abstraction} \label{sec:abs}
In computer science, an abstraction can be thought of as a model of a concrete artifact that retains the artifact's essential properties while eliding details of its implementation~\cite{abstraction2}.  Abstraction is used heavily across the interface of components to reduce the complexity of interactions across those components. Thus, it is quite natural that abstraction is heavily used at the interface between hardware and software, one of the most complex interfaces in modern computing. Some key properties of an abstraction, which is the interface created when one abstracts a concrete implementation, are 1) the attenuation of information and 2) the creation of new semantics.  For example, consider the implementation of modern processor caches:
\begin{enumerate}
    \item Attenuation of information: Explicit access to and control of cache memories is hidden, only revealing faster code execution.
    \item New semantics: Caches expose the abstraction of a cache line, the size of the cache memory, a mapping function between cache lines and memory and certain instructions (e.g., \texttt{WBINVD/PREFETCH/CLFLUSH}), which influence cache behavior.
\end{enumerate}
From this example, we can see that abstraction both hides and exposes information. Depending on the components involved, what 
can be hidden is the existence of resources or metadata. Other examples of hardware-software abstractions include the processor ISA, which exposes the semantics of an instruction for software to utilize hardware resources, but hides implementation details such as pipelining, out-of-order-execution and speculation.

We distinguish abstraction from isolation because while both reduce the flow of information, \textit{total isolation} as classically defined by Lampson~\cite{lampson1973a} means that there is no information flow at all, while abstraction merely attenuates it, still allowing information to flow over the new semantics it defines.  Similarly, virtualization is a particular instance of abstraction, which seeks to maintain one abstraction with a vastly different implementation (i.e. backing RAM memories with disk).  \looseness=-1

The security concern with abstraction is that as hardware has become a critical piece of security for software, hardware abstractions have been thrust into a role for which they were not explicitly designed, which is to provide access control between mutually distrustful, actively malicious entities. Failures of these abstractions in this role can have disastrous consequences.  For example, the Meltdown/Spectre vulnerabilities can be attributed to a failure in the ISA abstraction to properly restrict memory accesses by one process to another (or to the kernel), which we discuss in Section~\ref{sec:under}.  In another example, the hiding of information via the SMM abstraction key information about the SMM region of memory from flowing to reference monitors in the chipset and CPU caches, allowing unauthorized access to SMM state, which we discuss in Section~\ref{sec:over}.  From these examples, we can see that the correct design of hardware abstractions is crucial to providing trustworthy execution environments, and it is through abstraction, that we will analyze and model TEEs in this paper.

\subsection{Execution Environments} \label{sec:ees} \label{sec:caps}

While the ISA abstraction describes the abstraction over which software can use hardware resources, the more general interface that describes all interactions between software and hardware is the Execution Environment (EE), which includes not just the ISA, but other abstractions, such as concurrency and level of confinement.   

We call an EE that implements trustworthy execution a TEE.  What is trustworthy execution?  Trustworthy execution is an execution system whose goal is the correct execution of a task (as opposed to endpoint security, which concerns the whole computing device).  As such, the goal of trustworthy execution can be subdivided into two sub-goals:
\begin{itemize}
    \item Initial integrity: If the task is started wrong, no correct execution can be expected. Related to this, one TEE may invoke another TEE and propagate trust to it using a \emph{chain of trust}. 
    \item Run-time security: Once started, the task is subject to both attacks from the outside and misbehavior of internal code.  Such attacks may seek to directly subvert the task or indirectly exploit a defect in the task to subvert it.  If such attacks are successful, then the task will not execute correctly.
\end{itemize}
Several properties of TEEs are relevant to their ability to provide trustworthy execution.  As such, these properties can form a basis for evaluating and comparing TEEs to see if they are ``fit for purpose''.  The following are major security properties we consider for TEEs:
\begin{enumerate}
    \item \emph{Initial integrity assurance}. This describes how the TEE provides initial integrity, and in general, comes in one of two forms: static image integrity and launch integrity. The former checks only the image at installation (or update) time, applying mostly to firmware update mechanisms,
    while the latter checks both the image and initial inputs right before execution.
    \item \emph{Addressing/memory protection}. This describes both what memory the TEE can access, and whether its memory can be accessed by other EEs.  This determines the advantage/disadvantage of a TEE, i.e., if defense code is hosted inside, to what extent it can be protected from being manipulated (integrity) or even seen (confidentiality) from outside, and at the same time, what ability it has to monitor or control the execution of other EEs.
    \item \emph{Scope}. This determines whether the TEE is per-logical processor (LP), per-core, per-CPU or per-system.
    For example, it is possible to see whether a task is in an SGX enclave or not at the granularity of an LP, as the CREG  \texttt{CR\_ENCLAVE\_MODE}  which indicates this exists for each  LP~\cite{sgxexplained}. In contrast, TrustZone is per-system as a single \texttt{NS} bit, which is propagated from the main bus to the APB (peripheral) bridge~\cite{tzexplained}, exists for the entire system.

    \item \emph{Developer access}.  This describes whether the TEE is designed to run third party, possibly untrusted or malicious code.
\end{enumerate}
Note that we explicitly exclude the ISA-defined notion of privilege-level (i.e. the current privilege level (CPL) or ``Ring level'' in x86) from these properties.  This is because, with regards to security, the traditional hierarchy of ISA privilege has become less relevant, as they are not absolute (i.e. Ring 0 provides different protection levels depending on whether it is between a kernel and a user process, or a kernel and a VMM), and it is no longer a strict hierarchy (SGX enclaves run at Ring 3 but still resistant to Ring-0 code and ARM privileges are orthogonal to the secure/non-secure status).  

There are a number of ISA extensions that serve to hide or restrict information flow, but do not form a complete TEE as they do not provide all the abstractions necessary to execute code.  For example,  Intel SMEP/SMAP~\cite{ia32-64} (or ARM PXN~\cite{arm}) can only be used to restrict the addressing capabilities of privileged code, while  Intel MKTME~\cite{mktme} (cf. TME) lacks proper addressing isolation from outside code they do not protect.  Thus, these can be viewed as extensions to existing EEs.  

\begin{table*}
\centering
\resizebox{0.9\textwidth}{!}{%
\begin{tabular}{|@{\hskip3pt}c@{\hskip3pt}|c|c|c|c|@{\hskip3pt}l@{\hskip3pt}|c|}
\cline{1-7}
\begin{tabular}[c]{@{}c@{}} \end{tabular} &
\begin{tabular}[c]{@{}c@{}}TEE\end{tabular} & Type &
\begin{tabular}[c]{@{}c@{}}Scope\end{tabular}& 
\begin{tabular}[c]{@{}c@{}}Init integrity\end{tabular} &
\multicolumn{1}{c|}{Accessible by} &
Dev \\ \hline
\ding{182} & \multicolumn{1}{c|}{Microcode} & F 
& CPU & Launch & {\footnotesize \begin{tabular}[b]{@{}c@{}}\underline{\hspace{50pt}}\end{tabular}} & \\ \cline{1-6} 
\ding{183} & \multicolumn{1}{c|}{ME / PSP (-3)} & F 
& System & Static  &  {\footnotesize \begin{tabular}[b]{@{}c@{}}\underline{\hspace{50pt}}\end{tabular}} & \\ \cline{1-6} 
\ding{184} & \multicolumn{1}{c|}{S3 Boot Script\footnotemark} & F 
& System & Static & {\footnotesize \begin{tabular}[b]{@{}c@{}}\underline{\ding{182}\ding{183}\ding{185}\ding{186}\ding{187}\ding{188}\ding{189}\hspace{6pt}}\end{tabular}} & \\ \cline{1-6} 
\ding{185} & \multicolumn{1}{c|}{SMM (-2)} & F 
& CPU & Static & {\footnotesize \begin{tabular}[b]{@{}c@{}}\underline{\ding{182} \hspace{41pt}}\end{tabular}} &  \multirow{-4}{*}{N} \\ \hline 
\ding{186} & \multicolumn{1}{c|}{TXT / SVM} & A 
& System & Launch & {\footnotesize \begin{tabular}[b]{@{}c@{}}\underline{\ding{182}\ding{184}\ding{185}\hspace{31pt}}\end{tabular}} & \\ \cline{1-6} 
\ding{187} & \multicolumn{1}{c|}{VMM (-1)} & C 
& System & X & {\footnotesize \begin{tabular}[b]{@{}c@{}}\underline{\ding{182}\ding{183}\ding{184}\ding{185}\ding{186}\hspace{19pt}}\end{tabular}} &\\ \cline{1-6} 
\ding{188} & \multicolumn{1}{c|}{SEV-ES (0)} & A 
& N/A & Launch & {\footnotesize \begin{tabular}[b]{@{}c@{}}\underline{\ding{182}\ding{183} \hspace{35pt}}\end{tabular}} &\\ \cline{1-6} 
\ding{189} & \multicolumn{1}{c|}{OS (0)} & C 
& N/A & X & {\footnotesize \begin{tabular}[b]{@{}c@{}}\underline{\ding{182}\ding{183}\ding{184}\ding{185}\ding{186}\ding{187}\hspace{12pt}}\end{tabular}} &\\ \cline{1-6} 
\ding{190} & \multicolumn{1}{c|}{Application (3)} & C 
& LP & X & {\footnotesize \begin{tabular}[b]{@{}c@{}}\underline{\ding{182}\ding{183}\ding{184}\ding{185}\ding{186}\ding{187}\ding{188}\ding{189}}\end{tabular}} & \\ \cline{1-6} 
\ding{191} & \multicolumn{1}{c|}{SGX (3)} & A 
& LP & Launch & {\footnotesize \begin{tabular}[b]{@{}c@{}}\underline{\ding{182}\hspace{45pt}}\end{tabular}} & \multirow{-6}{*}{Y}  \\ \hline
\end{tabular}}
\vspace{10pt}
\caption{Properties of example execution environments on x86. \textbf{C}=Chained TEEs, \textbf{F}=Firmware TEEs, and
\textbf{A}=Attested TEEs.}
\label{tab:properties}
\vspace{-15pt}
\end{table*}

\section{Trustworthy Execution with TEEs} \label{sec:anchor}

In this section, we examine several well known TEEs and formally categorize them into Attested Trusted Execution Environments, firmware TEEs and Chained TEEs.  The TEEs we examine and their properties are summarized in Table~\ref{tab:properties}.  The ``Type'' column indicates which of the 3 categories each TEE belongs to, and the other 4 columns correspond to the 4 properties of TEEs discussed in Section~\ref{sec:ees}. The ``Accessible By'' column gives addressing/memory protection as a relationship between the TEEs by indicating which TEE can access which ones according to their intended design (i.e. not taking into account vulnerabilities).  The access here refers to read/write access to the execution memory of the TEE in question, and for this reason, we include only the x86 TEEs for comparability (e.g., excluding ARM TrustZone).

\subsection{Attested Trusted Execution Environments} \label{sec:tee}

We define Attested Trusted Execution Environments to include environments that are traditionally thought of as TEEs, such as SGX, TXT, or TrustZone.  Today's processor architectures usually support one or multiple attested TEEs (even on MCU-like platforms, e.g., TrustZone profile M~\cite{nuvoton,nordic}). The defining features that separate attested TEEs from other TEEs are that attested TEEs provide launch initial integrity and attestation capabilities.


\subhead{Initial Integrity}
Attested TEEs implement launch integrity through explicit measurements upon code loading and data sealing.  Moreover, the code loading process also measures the integrity of the environment of the attested TEE, including any other software loaded up to that point.  TXT/SVM achieves this by collaborating with the TPM chip (firmware-based, integrated or discrete) as secure storage. The program (including the SINIT module) being loaded is first measured by the CPU, and the measurement is stored in TPM's volatile memory (Platform Configuration Registers) in the form of hash values. If the hash values do not match with the preset values in TPM's non-volatile memory (policies in NVRAM indices), execution will be aborted. SGX has this functionality implemented as part of the microarchitecture extension (i.e., inside the CPU), without relying on TPM (in the \texttt{MRENCLAVE} field of the protected memory instead, but non-volatile secrets reside in the SPI chip on the motherboard).   Data sealing ensures that a piece of data can only be retrieved on a specific platform when a specific program is running. This is implemented by encrypting and decrypting the sealed data  with a key derived from secrets in the hardware and program measurements.

\subhead{Memory Protection} 
Attested TEEs also have stronger memory protection. The way memory protection is implemented depends on whether the attested TEE is exclusive or concurrent.  A concurrent TEE coexists with the unprotected portion of the system (e.g., in a time-sliced fashion), while an exclusive TEE preempts other code and then destroys the execution context before allowing other code to run.  

Concurrent attested TEEs, such as SGX, have special processor mechanisms to provide strong isolation from other EEs, such as both privileged and unprivileged EEs on the same system.  SGX has its enclave memory allocated in the Enclave Page Cache (EPC) which is part of the Processor Reserved Memory range (PRM). The PRM's protection is enforced by the CPU against access from any other software (including SMM) and DMA. A unique feature of SGX is that the enclave memory is fully encrypted when exposed outside the CPU with the MEE (Memory Encryption Engine, part of the CPU uncore), immune to various (physical) memory attacks, e.g., the cold-boot attack~\cite{coldboot}. Thus even in the case of exposure from the EPC (e.g., when EPC pages are evicted into regular DRAM), memory content is only seen as ciphertext.

In contrast, exclusive TEEs do not need to defend against concurrent software threats and as a result, for these Attested TEEs, the main focus of memory protection is to defend against DMA access from peripherals and physical memory attacks. An Attested TEE such as TXT relies on Intel's IOMMU technology VT-d to protect its MLE from being accessed via DMA, by including the memory ranges in the DMA Protected Range (DPR) and Protected Memory Regions (PMRs)~\cite{txtmle}. SVM also has AMD's Device Exclusion Vector (DEV) support~\cite{amd64} for the same purpose. Both TXT and SVM are vulnerable to the cold-boot attack and accessible by SMM, unlike SGX.
This can be an example of low-privilege TEEs having stronger protection than perceived from their assigned privilege.

\footnotetext{We use the S3 boot script as an example to demonstrate typical protection level for BIOS/UEFI firmware.}

\subhead{Scope}
Attested TEEs vary in scope.  Many of the earlier attested TEEs, such as TXT, SVM and TrustZone encompassed the entire system, while SGX is per logical processor.  This reflects SGX's goal to be lighter-weight while the other attested TEEs were seen to be an entire virtual machine, complete with its own OS.  However, system-wide TEEs, especially TrustZone, take advantage of their system-wide property by being able to have trusted and possibly exclusive access to hardware peripherals.  Most system-wide TEEs are exclusive, in that they cannot execute concurrently with another OS.  However,  TrustZone~\cite{tz,tzexplained} (the traditional profile A) is an exception, thanks to its I/O partitioning capability. With the TrustZone Address Space Controller (TZASC)~\cite{tzexplained}, memory-mapped devices can be dynamically partitioned~\cite{secloak}, e.g., a part of the screen dedicated to the secure world, allowing concurrent access by both the TrustZone OS and an untrusted-OS running outside TrustZone. We note that in the context of Attested TEEs, system-wide TEEs 
are thought of as \textit{privileged TEEs} (pTEE), because they can run privileged code, while local-process TEEs, such as SGX, are thought of as \textit{unprivileged TEEs} (uTEEs).  

\subhead{Developer Access} 
Universally, Attested TEEs are meant to be open environments that allow arbitrary developer code to be executed in them.  As such, Attested TEEs also provide attestation, where the Attested TEE can sign the code measurements taken at launch with a trusted key to assert to another party the identity of the code executing within the Attested TEE.  For system-wide Attested TEEs, this comes in the form of remote attestation, as it attests the identity of the entire system to the remote party.  LP-scoped Attested TEEs like SGX, are capable of both local attestation to other code on the same system, as well as remote attestation to code on other systems.  

\subsection{Firmware Trusted Execution Environments} \label{sec:fw} \label{sec:rotcaps}

Firmware is software that implements functionality that is logically part of the hardware.  As a result, while it is software, it is implicitly (axiomatically) trusted just like real hardware.  Unlike  Attested TEEs which can be considered alternative EEs that implement trustworthy execution for software that requires it,  firmware is part of the trusted computing base (TCB) of all software, as it is responsible for critical system operations. 

Firmware exists in various components of a computer system, which has an effect on where the firmware  executes. System firmware (sFW) executes on the main CPU. This generally refers to BIOS/UEFI firmware, which performs the early (but complex) initialization before bootloaders~\cite{loaders} and the OS. However, sFW is not confined to only boot time---SMI handlers (in System Management Mode) and UEFI Boot Scripts (upon S3 wakeup), continue to run even after system boot is complete.  sFW also includes CPU firmware, which is used to implement various instructions that may be called by software.  Chipset firmware (cFW) executes on a separate dedicated processor, often microcontrollers in a system's I/O subsystem, or in co-processors on the system board or system-on-chip (SoC).   Examples of cFW include the firmware for the Intel Management Engine (ME), AMD Secure Processor (previously PSP) and the baseband processor on  mobile phones.  Finally, Device firmware (dFW) executes on a dedicated processor as part of a peripheral.  The distinction from cFW here is that these devices are even more isolated from the host CPU, often accessed over a well-defined bus interface, which narrows the possible interactions between the firmware and untrusted code on the host CPU, e.g., SATA, PCI Express and USB.  An example of dFW could be the firmware on a hard drive or network card.  Another distinction is that the sFW and cFW are usually stored in non-volatile RAM that is shared with other sFW or cFW, or may even be loaded off disk and inserted by the BIOS or OS.  On the other hand, dFW is almost universally stored on the peripheral device.  

There are a large number of firmware TEEs with diverse properties, and we do not discuss each for the sake of space, but highlight some notable firmware TEEs for each property.

\subhead{Initial Integrity}
One property common to all firmware TEEs is that they provide static integrity as opposed to launch integrity like attested TEEs.  This means that the integrity of their code is only checked when it is updated and not when they are executed.  There are a few exceptions to this rule.  Both S-CRTM (Static Code Root of Trust for Measurement)~\cite{glossary} and  UEFI Secure Boot~\cite{secureboot} will measure certain firmware such as Option ROMs (which are device-specific firmware executed during boot to initialize a peripheral) during boot.  S-CRTM stores these measurements in the PCRs on a TPM, making them available for remote attestation.  UEFI secure boot, on the other hand, verifies the ROMs against some specified policy.  Finally, CPU microcode is loaded from the BIOS and verified on each boot.

\subhead{Memory Protection}
Each firmware TEE has its own custom memory protection mechanism.  The mechanisms may also permit access by some TEEs while denying access by others.   Table~\ref{tab:properties} lists which TEEs are able to access other TEEs.  
For instance, CPU microcode is able to access all other TEEs as it helps enforce the ISA abstraction, while its own internal state (e.g., the SRAM) is invisible (abstracted away) to the rest of the system. Intel ME as chipset firmware is not accessible by other TEEs (with shared memory encrypted). For management purposes, it has access to most part of the system with the exception of SMM and TXT. SMRAM is protected by the ISA (see Section~\ref{sec:under} for attacks) and only exposed to microcode. The S3 boot script is a special case in firmware TEEs but represents certain other UEFI modules: it is granted access to the whole system memory and I/O (to resume the pre-sleep machine state) but where it resides is open to any privileged code (if UEFI LockBox is not implemented).

\subhead{Scope}
In view of the special nature of firmware TEEs, we discuss their scope as follows. CPU Microcode runs underneath (and helps create) the ISA abstraction. Its scope can be considered the entire CPU. Intel ME (or similar chipset technologies) has one instance and coordinates the whole computer system, regardless of other code on the main processor(s). Therefore we consider ME's scope to be System. Certain UEFI/BIOS modules handle specific system stages (e.g., self-check, power state switching, and device initialization), during the inactivity of other tasks on the main processor(s). Hence the S3 boot script's scope is assigned System. 

Proper scope can ensure a firmware TEE, if used to host defense code, of sufficient coverage for enforcement. For example, mechanisms based on Intel ME are able to overwatch code of various privileges on different processors in the system. On other hand, mechanisms using SMM in a multi-processor environment must consider interaction with other processors, as SMI processing is per-processor~\cite{smm2}.

\subhead{Developer Access}
Another distinguishing property between firmware TEEs and Attested TEEs is that Attested TEEs were inherently designed to host developer code, while firmware TEEs are designed for the exclusive use of code belonging to the device or platform manufacturer, and as such, are not open execution environments.  As a result, firmware TEEs lack features like remote attestation, and even the specific instances where their measurements are included in attestations were bolted on as after-thoughts to increase the trustworthiness of those attestation by encompassing more of the TCB of the host.  There have been some academic proposals to inject code via non-officially supported methods into a firmware TEE to improve host security overall (in particular SMM).  We discuss these in Section~\ref{sec:ee}.

\subsection{Chained Trusted Execution Environments} \label{sec:howto}

Some EEs do not inherently provide trustworthy execution, mainly because they do not provide initial integrity and are not designed to provide run-time security.  However, such EEs can be turned into TEEs by chaining trust and adding secure functionality.  For example, kernel mode execution is not innately a TEE, but if we boot a verified secure OS (such as SEL4~\cite{KleinseL4formalverification2009} for example), and chain trust to it by attesting it using a lower-level TEE such as TXT or a TPM, then the kernel mode EE \textit{becomes} a TEE.  We call TEEs created in this manner ``chained TEEs''.  

\subhead{Initial Integrity}  The initial TEE to execute is called the \textit{Root of Trust} (RoT) as it is trusted by fiat.  It can then chain that trust to other EEs to make them into TEEs using the following sequence of rough steps:
\vspace{3pt}
\begin{enumerate}[\indent(a)]
    \item Establish a mechanism to check and protect the next EE.
    \item Ensure that the EE's coverage is sufficient for the intended task. 
    \item Transfer execution to the checked/protected EE and optionally repeat by going to (a).
\end{enumerate}
\vspace{3pt}
For example, an initial TEE (e.g., UEFI, as the Root-of-Trust TEE) boots the OS as the next TEE with Secure Boot or Boot Guard, and then the OS can check and run an appropriate anti-virus tool.  Most Attested TEEs can act as the Root-of-Trust TEEs. In ``late launch'' technologies (Intel TXT and AMD SVM), the privileged TEE backed by hardware directly bootstraps an OS/VMM without relying on UEFI/BIOS; furthermore, an unprivileged TEE (uTEE, e.g., Intel SGX) can securely bootstrap the ultimate user task skipping also the OS/VMM.  Note that in both cases above, legacy firmware still remains part of the TCB to a certain extent, but is not the Root-of-Trust TEE for the chain of trust establishment.

\subhead{Memory Protection, Scope and Developer Access}
Such chained TEEs are generally the VMM, OS or application code (if trust is chained from the privilege layers below).  As such, their memory protection and scope are that of the VMM, OS or application code depending on the program being executed and configuration.
We do not consider the scope of the OS (correspondingly the SEV/ES), as one OS may span processors/LPs but there can be multiple instances in the case of virtual machine guests (under a VMM).
In general, these systems derive memory protection through the processor MMU and are system-wide for the VMM, 
and limited to a logical processor for an application.  All are open EEs and support the execution of developer-specified code.

\begin{table}
\centering
\begin{threeparttable}
\caption{Distribution of EEs across different platforms (examples).}\label{tab:armx86z}
\begin{tabular}{clllc}
\hline
\multicolumn{3}{|l}{\textbf{\begin{tabular}[c]{@{}l@{}}$\Longleftarrow$ more integration\end{tabular}}}                                                                                                                                                                                                                & \multicolumn{2}{r|}{\textbf{\begin{tabular}[c]{@{}r@{}} more offloading/discreteness $\Longrightarrow$ \end{tabular}}}                                                                                     \\ \hline
\multicolumn{2}{|>{\columncolor[gray]{0.95}}c|}{\backslashbox{EE}{Arch}}                                                                                                         & \multicolumn{1}{c|}{\cellcolor[HTML]{EFEFEF} ARM}                                                          & \multicolumn{1}{c|}{\cellcolor[HTML]{EFEFEF} x86}                                                                                    & \multicolumn{1}{c|}{\cellcolor[HTML]{EFEFEF} System Z}                                                                                                                                                    \\ \hline
 \multicolumn{1}{|>{\columncolor[gray]{0.95}}c|}{\multirow{-1}{*}{dFW}} & \multicolumn{1}{>{\columncolor[gray]{0.95}}c|}{\begin{tabular}[c]{@{}c@{}}\scriptsize Stand-alone\\ \scriptsize entities\end{tabular}} & \multicolumn{1}{c|}{}                                                             & \multicolumn{1}{c|}{}                                                                                       & \multicolumn{1}{c|}{\begin{tabular}[c]{@{}c@{}}HMC\\ SE\\ CSS\\ CU\\ Director/Switch\end{tabular}}                                                                                                                                                                        \\ \hline
\multicolumn{1}{|>{\columncolor[gray]{0.95}}c|}{} & \multicolumn{1}{>{\columncolor[gray]{0.95}}c|}{\begin{tabular}[c]{@{}c@{}}\scriptsize Characterizable\\\scriptsize Processors\end{tabular}}                                                               & \multicolumn{1}{c|}{} & \multicolumn{1}{c|}{} & \multicolumn{1}{c|}{\begin{tabular}[c]{@{}c@{}}zIIP\\IFL\\ICF\end{tabular}}                                                                                                                                                            \\ \hhline{~----} 
  \multicolumn{1}{|>{\columncolor[gray]{0.95}}c|}{\multirow{1}{*}{cFW}}                   & \multicolumn{1}{>{\columncolor[gray]{0.95}}c|}{\begin{tabular}[c]{@{}c@{}}\scriptsize Apps on\\\scriptsize co-processors\end{tabular}}                                                               & \multicolumn{1}{l|}{}                                                             & \multicolumn{1}{c|}{\begin{tabular}[c]{@{}c@{}}AMT\\PAVP\\fTPM\end{tabular}}                                                                                       & \multicolumn{1}{c|}{\begin{tabular}[c]{@{}c@{}}i390 code \\(on SAPs)\\CFCC \\(on ICF)\end{tabular}}                                                                                                                                                            
\\ \hhline{~----} 
\multicolumn{1}{|>{\columncolor[gray]{0.95}}c|}{}                     & \multicolumn{1}{>{\columncolor[gray]{0.95}}c|}{\scriptsize Co-processors}                                                               & \multicolumn{1}{c|}{\begin{tabular}[c]{@{}c@{}}Baseband\\ Apple SEP\\SCP/MCP\end{tabular}}                                                             & \multicolumn{1}{c|}{\begin{tabular}[c]{@{}c@{}}Intel ME\\ AMD PSP\end{tabular}}                                                                                       & \multicolumn{1}{c|}{\begin{tabular}[c]{@{}c@{}}SAP\\IFP\end{tabular}}                                                                                                                                                      
\\ \hline
\multicolumn{2}{|c|}{\cellcolor[HTML]{EFEFEF} sFW}                                                                             & \multicolumn{1}{c|}{Various}                                                             & \multicolumn{1}{c|}{SMM}                                                                                       & \multicolumn{1}{c|}{PR/SM}                                                                                                                                                             \\ \hline
\multicolumn{2}{|>{\columncolor[gray]{0.95}}c|}{\begin{tabular}[c]{@{}c@{}}Privilege\\system\end{tabular}}                                                                          & \multicolumn{1}{c|}{EL/PL0--3}                                                             & \multicolumn{1}{c|}{Ring0--3}                                                                                       & \multicolumn{1}{c|}{16 keys x 2 states}                                                                                                                                                                    \\ \hline
\multicolumn{2}{|c|}{\cellcolor[HTML]{EFEFEF} pTEE}                                                                                & \multicolumn{1}{c|}{TrustZone}                                                    & \multicolumn{1}{c|}{\begin{tabular}[c]{@{}c@{}}Intel TXT\\ AMD SVM\\AMD SEV\end{tabular}}                            & \multicolumn{1}{c|}{}                                                                                                                                                                \\ \hline
\multicolumn{2}{|c|}{\cellcolor[HTML]{EFEFEF} uTEE}                                                                               & \multicolumn{1}{c|}{Planned\footnotemark}                                                             & \multicolumn{1}{c|}{Intel SGX}                                                                              & \multicolumn{1}{l|}{}                                                                                                                                                             \\ \hline
\end{tabular}
    \begin{tablenotes}
      \small
      \item Refer to Appendix for the glossary and further explanation.
    \end{tablenotes}
  \end{threeparttable}
\vspace{-15pt}

\end{table}

\footnotetext{As of this writing, no official documentation or public information is available for ARM Bowmore, a technology for isolating individual workloads. We note it here for completeness.}

\subsection{TEEs across architectures}
In light of the importance of TEEs in establishing (the chain of) trust for a computing platform, we review the presence and positioning of TEEs across mainstream architectures (refer to Table~\ref{tab:armx86z}). ARM (most mobile platforms), x86 (most PCs and servers) and mainframes (aka, System Z, e.g., in data centers) are included.

\subhead{Observations}
We notice that the positioning of TEEs reflects the purpose of the corresponding platform.
ARM/x86 aim to be tightly integrated, migrating more towards few co-processors that are either on the board or even on a ``system on chip'' ; while System Z aims to be highly modular, and contains many discrete components.
Compared to commodity platforms like ARM and x86, System Z trades-off cost for greater security, availability, reliability and performance. To reduce the chance of single point of failure and to offload processing to task-specific components, mainframes like System Z have the following distinctions in terms of TEEs:

\begin{itemize}
    \item Offloading to co-processors. System Z is a coprocessor-rich platform~\cite{zhardware}. There are basically two types of co-processors: 
    \begin{inparaenum}
        \item Unconfigured generic processors. Physical processor units (PUs) are shipped generically and must be configured (\emph{characterized}~\cite{zconcepts}) by the customer for a purpose, such as zIIP (for Java and database workloads) and IFL (for Linux workloads).
        \item Configured processors. Certain PUs are configured with a default purpose by the vendor (corresponding to co-processors on x86/ARM), e.g., IFP runs firmware that is specific to certain PCIe features.
    \end{inparaenum}
    These co-processors contribute to the performance and availability of System Z.
    \item Stand-alone components. Components that are usually integrated on PCs take the form of one or multiple stand-alone devices on System Z, e.g., BMC (on-board) vs. the Hardware Management Console and the Support Element (a separate laptop). 
    \item Full-stack implementation. Furthermore, compared to regular containers (LXC/Docker) which share the underlying abstraction layers up to the OS, IBM Secure Service Container (SSC~\cite{ssc}) has its dedicated hardware, firmware and OS in a physical box.
    \item Fine-grained privilege levels. There are 16 storage access keys (combined with the 2 states/privileges), assigned to different workloads individually. This forms an architectural support for fine-grained privileges.
\end{itemize}

To maximize security out of co-located entities (as a result of integration for portability), ARM/x86 tends to have rich support for Attested TEEs, which are intended for isolation (enhancing abstraction) in a shared computing environment. 

\section{Abstraction Underdone} \label{sec:under}
As an important form of trust establishment, TEEs are expected to abstract information away sufficiently from designated entities (hence achieving protection). However, even if the hardware implementation complies with the abstraction specification (e.g., an ISA), information can still be leaked in one way or another.  We now examine how insufficient abstraction opens the door for attacks.

\subhead{Side channels}
The term side-channel attack originated from cryptography~\cite{timing}. It has been used to refer to secret extraction from unintended channels such as timing, power, electromagnetic and acoustic channels.
We generalize the side channel and define it as an unauthorized communication channel caused by implementation details that are not specified by the abstraction specification, such as \emph{algorithm, protocol, architecture and interface standard}.

A TEE may have been properly implemented regarding what should be hidden and what should be exposed, but it has been shown in multiple incidents that the intended abstraction is insufficient.

\subhead{Synchronousness}
As side channels imply the presence of another entity (the adversary), whether that entity simultaneously and actively extracts information from the victim code determines the synchronousness.

Exclusive attested TEEs (with the scope of System) are naturally immune to synchronous side-channel attacks, as they do not execute in the presence of any other software. We do not consider physical side channels here (e.g., power analysis attacks~\cite{ohmslaw}). This is also reflected in the fact that almost all concurrent TEEs suffer from side-channel attacks to a certain extent (see Section~\ref{sec:teesec}).

As for exclusive firmware TEEs, direct run-time side-channel attacks are not applicable. However, attacks caused by improper protection of their memory (when inactive) are common, as exemplified by the attacks discussed in Section~\ref{sec:fwees}.

\subsection{Micro-architectural side channels}
The processor architecture (e.g., the ISA) forms an abstraction layer between software and hardware. When the ISA specification is complied with, side channels caused by the underlying implementation are micro-architectural. 

Most micro-architectural side channels are timing-based, i.e., extracting information by measuring temporal characteristics of architectural artifacts. This is because while the ISA abstraction specifies instructions and visible architectural resources like registers, it does not provide an abstraction for timing, instead allowing the underlying implementation to determine operational delays in the interest of maximizing performance.

There are two aspects to such side channels:
\begin{itemize}
    \item Information extraction. Certain (mis)behaviors effectively convert micro-architectural data to architectural, to be extracted. Two subcategories exist:
    \begin{inparaenum}
        \item Memory content. The wide variety of cache-based side channels exploit the sharing of different cache levels to derive secrets from cache access (hits or misses), as represented by Prime+Probe~\cite{primeprobe} and Flush+Reload~\cite{flushreload}. In addition to regular cache, the Translation Lookaside Buffer (TLB) as the cache for the Memory Management Unit (MMU) also leaks information (e.g., TLBleed~\cite{tlbleed}). Such indicates the abstraction flaw of cache memory beyond the ISA.
        \item Execution metadata. The CPU's Execution Unit, if not properly abstracted, can leak diverse metadata about ongoing execution. PortSmash~\cite{portsmash} can time the contention latency of execution engine ports, and infer instruction traces based on port assignment difference. Nemesis~\cite{nemesis} learns the current instruction in execution according to interrupt handling latency.
        Note that memory content can be further extracted based on the execution metadata.
    \end{inparaenum}
    \item Channel control. Incomplete abstraction can also enable the attacker to better control what is leaked over the side channel. For example, vulnerabilities like Meltdown~\cite{meltdown} and Spectre~\cite{spectre} exploit the incomplete abstraction of speculative execution to select what information is leaked with greater probability over cache-based channels (for the actual information extraction).
\end{itemize}

\subsection{(In)security of concurrent Attested TEEs} \label{sec:teesec}
Most attested TEE side channels are also micro-architectural side channels with certain adaptation specific to the Attested TEE.

SGX heavily suffers from side-channel attacks due to its concurrent but unprivileged nature, ranging from branch shadowing~\cite{branchshadow}, cache attacks~\cite{sgxcache}, SgxPectre~\cite{sgxpectre} to the fatal Foreshadow attack~\cite{foreshadow}. There exist also mature SGX-specific compromise tools to facilitate attacks, e.g., SGX-Step~\cite{sgxstep} that single-steps enclave code.
All in all, regular side channels might be alleviated by programmer diligence and involving another root anchor (e.g., Intel TSX~\cite{tsgx,dejavu}), and Foreshadow is patchable (although its long-term influence is an open question).  TrustZone, as a concurrent attested TEE is also vulnerable despite its ability to run privileged OS code.  Side channels have been identified as in TruSpy~\cite{truspy} and TruSense~\cite{trusense}. Nevertheless, AutoLock~\cite{autolock} demonstrates that they might be more difficult than expected.  SEV, which  enhances VM guests with memory encryption, is vulnerable to secret extraction attacks~\cite{severed,sevattack2}, due to a malicious VMM being able to execute concurrently with a victim VM.

\subsection{Abstraction for firmware TEEs} \label{sec:fwees}
Compared to the abstractions used to construct Attested TEEs, firmware TEEs have comparatively simple abstractions as they tend to execute exclusively and tend to have very little interaction with application code in the host CPU.  Instead, the abstraction failures tend to lead to lapses in access control that allow adversaries to corrupt firmware TEEs, leading to compromises of the initial integrity of firmware TEEs.  We survey several documented instances of such lapses below.

\emph{System Management Mode (SMM)}: System Management Interrupt (SMI) handlers are stored in a RAM region called SMRAM, which is protected by the CPU at all times (as in the abstraction specification). The content of the SMRAM determines the initial integrity of the SMM TEE 
(after loaded from the SPI flash).

In early motherboards, access was possible from any kernel-privileged code, because previously the \texttt{D\_LOCK} bit in the SMRAM Control Register (\texttt{SMRAMC}) was not taken care of by the BIOS.\footnote{Once \texttt{D\_LOCK} is set, \texttt{SMRAMC} becomes read-only irreversibly until reboot, locking down \texttt{D\_OPEN} (which makes SMRAM visible.)}
Later, SMI handler compromise could take several forms (in addition to the SPI reflash attacks~\cite{biospersistent,biosattacking,biosdefeated}):
\begin{inparaenum}
\item through the memory reclaiming mechanism (e.g., intended for saving space wasted by MMIO)~\cite{smmpreventing} (patched on certain machines). Note that remapping is locked in Intel TXT mode~\cite{datasheet}.
\item cache poisoning~\cite{smmattack,smmreloaded}: access to improperly cached SMRAM content (fixed with the SMRR register).
\item SMM callout vulnerabilities~\cite{call_outside2}: SMI handler branches outside of SMRAM (fixable with \texttt{SMM\_Code\_Chk\_En}).
\item attacking argument passing to SMI handlers~\cite{smiarg}: tricking the SMI handler to overwrite SMRAM.
\end{inparaenum}
See Section~\ref{sec:lack} for attack details.
All these are to do with (previously) unspecified aspects between the SMM TEE and the rest of the system.

\emph{UEFI Boot Script} (which is run when the system wakes up from S3 sleep) can also be altered maliciously~\cite{uefibootscript}, thus allowing arbitrary code execution. The root cause is that the EFI variables or their copies (in this case a pointer to the Boot Script) are not properly protected. An attack on UEFI Secure Boot~\cite{securebootattack} was based on a similar approach (i.e., modifying an EFI variable storing the boot policy).

\emph{Microcode}: x86 microcode updates are initiated by writing to model-specific registers (MSRs) and accepted after certain cryptographic verification.\footnote{Microcode patches are not persistent and are reloaded during the early boot process (e.g., from CPUCODE.BIN in the SPI flash).} An early documented attempt was found in an anonymous report~\cite{microcode0} which showed an example of abstraction underdone that even if only vendor-verified updates are allowed, an attacker in control of this process can still choose to patch microcode lines that facilitate his attack (there are multiple slots available). 

\subhead{cFW TEEs}
Chipset FW TEEs, on the other hand, can suffer from inadequate abstraction both at run-time (synchronous) or when inactive (asynchronous), due to their concurrency on another processor while sharing the code storage with the main processor.

\emph{Intel Management Engine (ME)}: ME's intended functionality for full-control out-of-band management requires bulk data transfer capability (in addition to HECI for signalling or small amount of data) with the main processor.
Therefore, DMA is constantly active via a mechanism called UMA (Unified Memory Architecture), which is  used between the GPU and the CPU. Due to the limited memory space on the ME processor, it uses the UMA region (like stealing part of the host memory) as its execution RAM~\cite{mesecrets}. This opens up an attack vector. Researchers have already started their exploration since the early days of ME (see \cite{me} by Tereshkin and Wojtczuk). Basically, the approach was similar to that of the SMM compromise: remapping the ME UMA region (otherwise protected) to be accessible by the main CPU. Fortunately (or unfortunately for the defense-purpose community) Intel introduced UMA protection for both integrity and confidentiality (using encryption)~\cite{ruan}.

\subhead{dFW TEEs}
Contrary to sFW and cFW, dFW does not share the processor or the code storage, and is exposed to the main processor only through a (limited) peripheral interface. Therefore, dFW's security rests less on abstractions in the processor and more on those at their interface.

A recent analysis of the (in)security of today's Self-Encrypting Drives (SEDs) identified several vulnerabilities~\cite{sed-attacks} (similar to regular SSDs).  Worryingly, not all attacks require physical access--- software-only reflashing attacks using undocumented vendor-specific commands (VSCs)\footnote{These VSCs are like regular commands sent through the SATA or NVMe interface.} could potentially be performed after a privilege escalation.  The attacker's code would remain on the peripheral even if the operating system is wiped and re-installed.

Indeed, dFW has been a battlefront for  attacks for decades. Examples are not rare, e.g., hard-drive backdoor~\cite{harddrive}, network cards~\cite{nic} and video cards.
Zhang et al.\ proposed IOCheck~\cite{iocheck} using SMM to monitor and verify the integrity of the firmware of various devices.  Hendricks and van Doorn~\cite{shoringup} gave a very high-level description of how device firmware can be verified in a trustworthy manner. However, the implementation of their proposal remains an open problem, largely due to the heterogeneity of the devices.

\section{Abstraction Overdone} \label{sec:over}
From the discussion in Section~\ref{sec:under}, we see that abstractions can still allow information flow in unintended ways, resulting in side channels across shared (though logically separated) hardware resources  (e.g., processors sharing the same last-level cache, firmware TEEs sharing the same SPI flash chip, etc.).  Conversely, we also find that abstractions, in their goal to attenuate information flows, sometimes go too far and hide information that needs to flow.  In many cases, this results in a hardware reference monitor enforcing an incomplete policy due to incomplete information.  We call this ``Abstraction Overdone'', and analyze some examples of this phenomenon in this section.

\subsection{Insufficient information flow for enforcement} \label{sec:lack}

By design, the CPU maintains a number of registers (or state information in other forms) that are internal, and the chipset also has its own internal state, e.g., in the case of the PCH (Platform Controller Hub). Some of the previous attacks exploited such asymmetric information, e.g., different views between the CPU and the chipset. 

With Intel, there is a mechanism to reclaim the memory lost to MMIO below 4GB, exposing two registers \texttt{REMAPBASE} and \texttt{REMAPLIMIT} in the Memory Controller Hub (MCH, aka the NorthBridge)~\cite{datasheet}. At the same time, critical regions, such as the SMRAM region should not be remapped and exposed to any software running on the CPU.  However, the abstraction of SMM says that this mechanism should not be exposed to any components outside of the CPU (or even software on the CPU), and as a result, the chipset is not aware of this restriction (e.g., remap should be checked against \texttt{SMBASE}, a Model-Specific Register).  Rutkowska and Wojtczuk~\cite{smmpreventing} found that this omitted information flow allows SMRAM to be remapped using the reclaiming mechanism and made accessible to code on the CPU, which is a violation of SMM's security guarantees.   A strikingly similar vulnerability was found against the seemingly more powerful Intel ME, by remapping part of the Unified Memory Architecture (UMA) using the same reclaiming mechanism\footnote{ME requires UMA for run-time storage, due to the limited memory capacity of its own microcontroller.} (see \cite{me}). Even though later Intel introduced UMA protection so that this region is fully encrypted~\cite{meturnedoff}, it does not address the root cause, which is that abstractions prevent the reclaiming mechanism from checking remapping requests against some global list of sensitive memory ranges.  With AMD, there is still no exception: the Input Output Remap Registers (IORRs) can be used to achieve the same purpose~\cite{iorr}.

Caching is another abstraction where insufficient information flow has resulted in vulnerabilities. Code running on the CPU can determine whether/how memory regions can be cached, through the Memory-Type Range Registers (MTRRs).
However, since the MCH does not see more than what is specified by MTRRs, it is unaware of memory access restrictions similar to the reclaiming mechanism. For example, a cache poisoning attack  proposed by Duflot et al.\cite{smmreloaded} and again by  Wojtczuk and Rutkowska~\cite{smmattack}, found that an  adversary could modify the ``exposed'' SMI handler while it resided in the cache without accessing the SMRAM directly.\footnote{We see some indication~\cite{smmcachepatent1,smmcachepatent2} that Intel was also likely aware to some extent of these vulnerabilities before their public disclosure.} On the next invocation, the modified SMM handler will run.  Duflot et al.\ also discussed a more efficient scheme to make the attack persistent and not confined by the cached size. Both assumed that the original SMI handler did not flush caches before the \texttt{RSM} instruction. The fix for this problem is the System-Management Range Register SMRR~\cite{ia32-64}, which complements the MTRRs by allowing the chipset to specify memory ranges that should not be cached (and can only be modified while the processor is in SMM).

\begin{figure}[h]
\centering
	\includegraphics[width=0.5\textwidth]{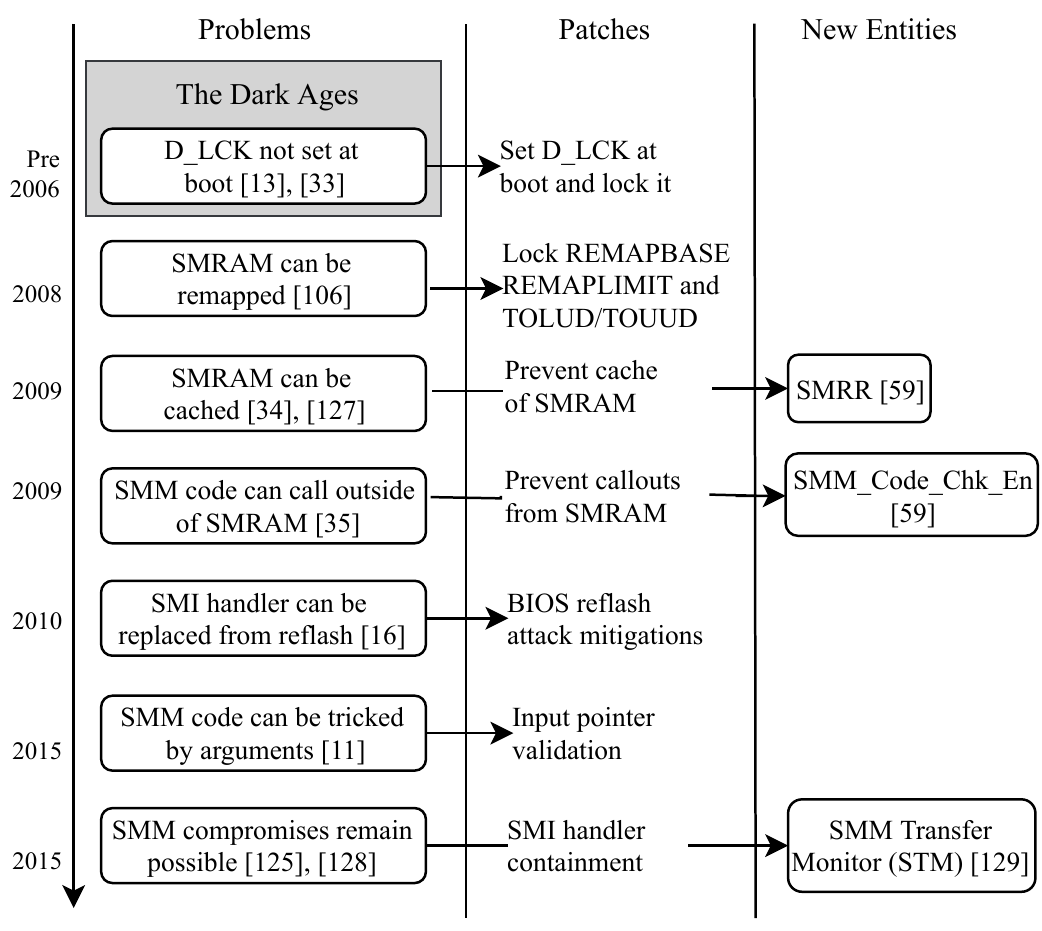}
	\vspace{-5pt}
	\caption{The arduous journey of SMM defences. 
	}
	\label{fig:smm} 
\end{figure}

If we examine how SMM attacks and defenses have evolved over the past decade (Figure~\ref{fig:smm}), we see that it has been an arms race. Namely, SMRAM is supposed to be protected, but what would be the defense vectors depends on what has been discovered by the community, gradually in an attack-driven manner.  While the MTRRs allowed information about the SMM memory range to cross the privileged-code EE abstraction to the OS kernel, the abstracted interface between the chipset and the CPU did not transmit this information until the SMRR was added.  In every instance, the root cause can be attributed to an overly strong abstraction preventing information about an access policy from flowing to the appropriate reference monitor.

\nocite{smm2006}

\nocite{smmpreventing} \nocite{d_lck2}

\nocite{smmreloaded} \nocite{smmattack} 
\nocite{ia32-64}

\nocite{duflot2010system} \nocite{call_outside2}

\nocite{bulygin2014summary}

\nocite{smiarg}

\nocite{uefi_attacking} \nocite{biosattacking}

\nocite{stm1}

\section{Proposed TEE-based security approaches} \label{sec:ee}
Defending against malicious code from outside (as defined in Section~\ref{sec:ees}) is usually the primary goal of security solutions (while avoiding internal code misbehavior can be the next step). This largely depends on 
TEEs that abstract away information/access from malicious code, e.g., in the form of memory protection or isolation. 
In this section, we look at proposals that use  TEEs to improve the security of systems.

\vspace{5pt}
\subhead{Usage 1: Exclusive pTEEs for initial integrity with a lightweight mechanism for run-time protection}
A number of solutions use privileged TEEs (pTEEs) to securely bootstrap the system into a good state, then hook critical operations together with metadata with a run-time mechanism. The rationale is as follows:
\begin{inparaenum}
\item pTEE's exclusiveness and load-time integrity measurement is used to bootstrap trust. 
\item The exclusiveness determines no monitored code can run in parallel, while switching back and forth with the monitored code imposes significant overhead, which is common for Attested TEEs. Note that hosting the whole solution/system inside the pTEE is infeasible either due to bloated TCB.
\item Therefore, for run-time protection, a common practice is to choose a hardened but lightweight mechanism. For example,
SMRAM's memory protection (inaccessible when SMM is not active) ensures continued integrity, and SMI's non-maskability enables it to preempt and monitor the rest of the system in a way that cannot be disabled.
\end{inparaenum}

\emph{Hypervisor integrity}: HyperSafe~\cite{hypersafe} makes use of tboot (which is actually based on Intel TXT) to bootstrap a solution for hypervisor integrity protection. It achieves a hardware-based memory lockdown for hypervisor pages by first protecting the pages with W$\oplus$X, then trapping any writes to page tables with the \texttt{WP} bit in \texttt{CR0}.
The enforcement logic (e.g., unlocking) is implemented in the page fault handler. Here the initial integrity enforced by TXT is critical for both the \texttt{WP} bit and other logic such as the page fault handler.

HyperSentry~\cite{hypersentry} proposes hypervisor-unaware integrity checks (to prevent a scrubbing attack where a compromised hypervisor hides the compromise from the monitor). The triggering logic is located in SMRAM as SMI handlers, but the core checking logic runs outside. To address the checking logic's lack of access to VM state information, the authors came up with a fallback mechanism with two consecutive SMIs that ensures
landing in the VMX root mode. The Intelligent Platform Management Interface (IPMI) is used as out-of-band signalling to trigger SMIs. 
HyperSentry assumes proper SMRAM protection from BIOS, in the form of trusted boot (S-CRTM) with TPM alone, instead of DRTM with TXT/SVM.

HyperGuard~\cite{smmpreventing} and HyperCheck~\cite{hypercheck} are two other examples of using SMM for hypervisor integrity. The difference is that HyperCheck outsources the core logic to the network (a remote machine), with the network card driver also in SMRAM; whereas HyperGuard collaborates with a chipset-based mechanism DeepWatch~\cite{deepwatch} (a trustlet in Intel ME). Both assume proper firmware protection (e.g., initial SMRAM integrity). 

\emph{Guest OS integrity}: CloudVisor~\cite{cloudvisor} protects the integrity of VM guests under the threat of compromised hypervisors. It also relies on TXT/SVM to ensure clean initialization. Then it provides a tiny security monitor using nested virtualization (equivalent to a lower-level VMM) to enforce isolation and protection of resources used by each VM guest.
Sensitive operations such as NPT/EPT faults, critical instructions and I/O
are trapped and examined by the tiny security monitor.

\vspace{5pt}
\subhead{Usage 2: Monitor directly from concurrent pTEE}
A concurrent pTEE can also be used to host an monitor, which directly monitors the OS/VMM without the help of an intermediary like SMM. Such a combined solution is not possible on x86 since the Attested TEEs on x86 are either unprivileged (uTEEs like SGX) and thus cannot monitor an OS/VMM, or switching is too expensive (exclusive pTEEs like TXT/SVM). 
The advantage of ARM TrustZone is that it is privileged and concurrent ( having low switch overhead as a result) at the same time. Examples of this usage are TZ-RKP~\cite{tzrkp} (which led to Samsung KNOX) and SPROBES~\cite{sprobes} to protect kernel integrity (in the normal world) by handling critical kernel events in the secure world. In both systems, kernel binary rewriting is needed to replace sensitive instructions with invocation of the Secure Monitor Call (SMC, a mode for switching to the secure world), which will invoke the TrustZone monitor.


\vspace{5pt}
\subhead{Usage 3: Containerization/isolation (TrustZone, SGX and SMM)}
Thanks to TEEs' addressing restrictions and memory protection (see Table~\ref{tab:properties}), they become a good candidate for containerization or isolation enforcement. While low-cost software sandboxes exist~\cite{NaCl,boxify}, a hardware-backed TEE can be ``root-secure'', meaning that they can withstand an adversary who has privileged code access (i.e. control of OS or hypervisor).

Considering that there currently exist no fine-grained (i.e., unprivileged) secure application environments in the normal world on ARM, Sun et al.\ propose TrustICE~\cite{trustice}. Multiple isolated computing environments (ICEs, similar to Intel SGX enclaves) can be created in the normal world, managed by a trusted domain controller (TDC) in the TrustZone secure world. 

In an effort to peel the host OS or hypervisor off the TCB, but accommodate secure workloads in parallel,
SICE~\cite{sice} employs SMM to create and manage enclave-like isolated computing environments (aka. ICEs).
It supports two modes: time-sharing (legacy host and ICEs run alternately) and multi-core (dedicated cores for ICEs). The latter takes advantage of SMM's per-core exclusiveness. Note that SICE only uses SMRAM as a shelter to store the ICEs (up to 4GB); SMM as a mode only prepares and enters the isolated environment, with the ICE code running in regular protected mode. As with the use of TrustZone, in SICE the legacy host (e.g., the hypervisor) is still required to add an interface that invokes an SMI to trigger SICE.

Scotch~\cite{scotch} combines a uTEE (i.e., SGX) and SMM to achieve reliable \emph{Hypervisor resource accounting}, using both SGX and SMM as isolation mechanisms. The strong preemptiveness of SMIs is used to forward all necessary events (interrupts and hypercalls) to SMM where secure accounting code runs and results can be communicated to and stored in the guest-side SGX space.

We can also consider lifting privileged code into uTEEs for finer granularity, as is done by Richter et al.~\cite{oscontained} in porting certain OS components to SGX. In addition to containerization for protection granularity, SGX's strengths over pTEEs are also important for the OS, e.g., immunity to SMM attacks and encryption against memory attacks. To showcase the benefits and feasibility, they have adapted the encryption module \texttt{dm-crypt} and move it to the SGX enclaves. The limited number of eligible components and huge performance loss may hinder its adoptability.

Apart from the TEEs above, researchers also pay attention to underused x86 privileges (currently only Ring 0 for kernel space and Ring 3 for userspace) to form intermediate protection for sensitive user-space tasks. LOTRx86~\cite{lotrx86} is proposed for such a purpose. It defines a new mode (PrivUser) run in Ring2-x32 and uses Ring 1 as a Gate mode in and out of the PrivUser mode. Sensitive per-application operations (e.g., operations related to memory safety) can be harnessed at a privilege higher than the rest of the application but still lower than the OS, hence ensuring mutual security.


\vspace{5pt}
\subhead{Usage 4: Secure user-machine interaction}
As mentioned in Section~\ref{sec:intro}, trustworthy execution also involves secure data exchange with the human user or peripherals, besides external malicious code and internal buggy code. If data input/output is intercepted, the execution logic's correctness alone is no longer leading to trustworthiness.

\emph{Exclusive pTEE without I/O partitioning}:
If the attested TEE cannot be assured that untrusted code does not have access to peripherals, it has to be exclusive to achieve secure user interaction.
Secure user input is usually considered together with what the user sees/perceives (i.e., secure output), UTP~\cite{utp} hooks pre-configured user data entry events/transactions, such as confirming an online purchase. It redirects the user to an attested TEE session (TXT/SVM reusing the Flicker~\cite{flicker} framework) to see a simplified display of the transaction, takes user input (confirmation) from the keyboard, and sends the attested two pieces of information to the server. The OS is resumed after such a session. In this case, both the physical keyboard and display are occupied by TXT/SVM because of its exclusiveness, hence considered secure. Very similarly, Bumpy~\cite{bumpy} also makes use of TXT/SVM with Flicker to protect user keyboard/mouse inputs. It involves more components for better usability and security: a USB interposer (an ARM board that could be later integrated to the keyboard/mouse) and a Trusted Monitor (a smartphone). A keystroke is encrypted by the USB interposer, processed and verified by the Flicker session, user-confirmed on the Trusted Monitor and sent to the server.

TrustLogin~\cite{trustlogin} employs SMM to secure
password-based login with a novel method of short-circuiting the OS, i.e., intercepting keyboard activities (source) and NIC packets (sink).
Considering trusted display alone, Yu et al.\ propose an attested TEE+GPU approach~\cite{trusteddisplay} that introduces a microhypervisor-managed \emph{GPU separation kernel} to serve the unmodified OS/applications and the secure applications at the same time. They employ XMHF~\cite{xmhf} as the microhypervisor with the TrustVisor~\cite{trustvisor} extension, which uses TXT/SVM for initial integrity. 

\emph{With I/O partitioning}:
On ARM platforms, secure user interaction seems to have attracted more attention, possibly due to the rich sensor environment and highly personal data storage.
As mentioned in Section~\ref{sec:tee}, TrustZone has I/O partitioning capability with its Protection Controller (TZPC) or TrustZone Address Space Controller (TZASC). This provides two important advantages:
\begin{inparaenum}
    \item It ensures exclusive resource access. Because of partitioning, even outside TrustZone, the normal world still cannot access the protected resource.
    \item The allocation of I/O resources between the secure and normal worlds can be changed dynamically.
\end{inparaenum}

\begin{itemize}
    \item User interaction. TruZ-Droid~\cite{truzdroid} proposes to move the sensitive UI interaction into the TrustZone secure world, while still maintaining the binding between the UI interaction and normal-world app code through indirect references (server-side). The user enters sensitive data and confirms transactions only with the Trusted Applications (TAs) in the secure world, discernible with a hardware Indicator LED. A similar approach is applied by TrustUI~\cite{trustui}, which also provides a trusted path between the user and the mobile device using input and display randomization. Both proposals rely on the TZPC.
    \item Peripheral/sensor management. SeCloak~\cite{secloak} is designed to securely and verifiably place the device in a user-approved state (on/off). It leverages both TZASC (as secure memory of the s-kernel) and Central Security Unit (CSU, a custom TZPC). 
\end{itemize}

It is obvious that since secure input/output concerns physical I/O operations (privileged in almost all architectures), the involved TEE/EE must also be privileged to perform any checks.


\section{Exposing More for Security}
\label{sec:misc}

TEEs assume a model mainly with external threats which are addressed by information hiding (abstraction).
TEE-based security solutions achieve trustworthy execution by ensuring that no code or data tampering from outside can slip through. However, except for certain simple programs that can be formally verified, software bugs almost become a destined byproduct in pursuit of higher performance and lower software development costs. If the programs are not written in type-safe languages~\cite{MILNER1978348}, attackers may feed malicious inputs to exploit some of the dangerous bugs to corrupt or subvert the programs~\cite{war-in-mem}. Another complicating factor is that software per se is not monolithic; user-level programs may load third party libraries, similar to operating systems loading drivers. The code available for inspection may only end up occupying a small portion at run time, and loading a buggy component may result in the entire program to be vulnerable. Consequently, TEEs may not be able to defend against such internal misbehavior.

What is worse, software misbehavior is hard to efficiently address in an either software-only or hardware-only manner alone. 
Due to the growing complexity of software, instrumented code introduces significant performance overhead, e.g., as in EffectiveSan~\cite{effectivesan} and WPBOUND~\cite{wpbound} (instrumentation is necessary to expose execution metadata). On the other hand, it is almost impossible for hardware alone to distinguish correct execution from misbehaved execution. Software semantics like buffer bound, data type, or dynamic allocation is hidden above the software-hardware interface (i.e., the ISA). This situation is very similar to the abstraction overdone as discussed in Section~\ref{sec:over}, which can be alleviated by exposing more across the interface (decreasing abstraction). 


Exposing more across the ISA interface effectively leads to software-hardware collaboration in addressing intra-EE misbehavior.  We note that many of these solutions are not TEE specific, and can be applied to any EE, trustworthy or not.  However, they can be applied to TEEs to significantly increase the assurance that the software in the TEE is free of internal misbehavior.  
Corresponding solutions usually fall into two categories based on the information flow direction:
\begin{enumerate}
    \item Hardware-to-software (execution metadata). 
    The hardware feature passively collects information from the execution trace of the monitored code, and sends it to a monitoring component to detect misbehavior. 
    \item Software-to-hardware (application semantics).
    The hardware feature learns application semantics from the monitored code through certain hints (explicitly like new instructions or registers or implicitly as in $\mu$CFI~\cite{ucfi}), and stops the execution when specified rules are violated. 

\end{enumerate}
Control/data-flow attacks and other memory safety problems have been systematically studied~\cite{war-in-mem} in the literature.
We do not repeat the discussion (e.g., the taxonomy of memory attacks) and consider them all as intra-EE misbehavior.

Software-hardware collaboration usually involves a monitoring component that assists the local hardware for policy/metadata management.
Its implementation can range from local software for simplicity, a remote server for flexibility, to dedicated hardware for performance.
For instance, LiteHAX~\cite{litehax} 
is a remote monitoring and attestation scheme against both control-flow and data-only attacks on embedded devices, which uses a remote computer for analysis. 
In the case of local software monitoring component, we assume proper isolation/protection from the monitored code, as is addressed by TEE-based approaches.

\subsubsection{Hardware-to-software}
Hardware passively collects execution metadata in collaboration with its monitoring component. Note that hardware involvement is required here as the monitored code is not trusted to provide such metadata and other software has no access to it.
As the monitored code does not need to cooperate, an advantage of the hardware-to-software approach is backward compatibility, allowing uninstrumented original code to be monitored.

\subhead{Coarse-grain control flow integrity with existing hardware}
The ISA typically provides dedicated instructions to perform indirect control flow change (e.g., \texttt{call} or \texttt{ret}), and thus hardware can infer control flow information directly, assuming that most programs use these instructions in the intended way. For example, the Last Branch Record (LBR) registers store a limited amount of trace information, as is used by ROPecker~\cite{ropecker} and kBouncer~\cite{kbouncer} against Return Oriented Programming (ROP) attacks. They rely on patterns of control flow change as the attack signature, as well as looking for suspicious short code sequences as an artifact of ROP. Similarly, the Intel Processor Tracing (PT) also provides activity traces but with more details and control.
GRIFFIN~\cite{griffin} uses Intel PT to enforce both forward-edge and backward-edge control-flow integrity. 

\subhead{Limitations of execution metadata without application semantics}
Metadata other than control flow is usually not retrievable with the existing ISA, e.g., data access information. While modified hardware can catch such metadata, it is still insufficient for data-oriented attack mitigation without application semantics. For example, ARMOR~\cite{armor} (an approach to ensure data accesses within the allocated ranges) 
cannot confine each access to the intended object due to lack of semantic information.

\subhead{Monitoring privileged software}.
Critical data structures in OS/VMM tend to be more deterministic for hardware to monitor without per-application semantics. Kernel integrity monitor solutions are typical of such type,
e.g., Copilot~\cite{copilot} (periodical entire memory scanning with a PCI card), Vigilare~\cite{vigilare} (bus traffic snooping), KI-Mon~\cite{ki-mon} (monitoring with a co-processor) and MGuard~\cite{mguard} (using both a modified DRAM module and a co-processor).

\subsubsection{Software-to-hardware}
While hardware-to-software abstraction reduction preserves backward compatibility, limited application semantics becomes the major hurdle when better coverage or precision is needed. 
Therefore, exposing more semantic information from software to hardware becomes helpful. Instead of directly from the monitored code, this is usually done by 
a dedicated software component (e.g., a compiler) instrumenting the monitored code to expose its semantics (although programmer annotation is also seen).
The ISA is augmented with new instructions and/or registers.

\subhead{Precise control flow integrity with application semantics}
With valid code pointers marked by the monitored code, misbehaving execution that jumps to an unexpected address can be more easily identified.
Intel CET~\cite{cet} 
is a complete CFI defense for all privilege levels. For forward-edge protection, CET introduces a new instruction \texttt{ENDBRANCH} to mark valid indirect control transfer destinations, and the processor uses state machines to ensure proper indirect branch landing. A traditional shadow stack is used for backward-edge protection.
HAFIX~\cite{hafix} 
is based on the observation that a function cannot return if it is not called yet, targeting coarse-grained back-edge control flow protection.
Each function is assigned a unique label (marking start and end) and tracked with dedicated label state memory.
New instructions are introduced to update the label state.

\subhead{Memory access control using an application policy}

In hardware-to-software solutions to data-oriented attacks, memory requests are coarsely categorized as instruction fetch, load, and store, and bad requests can slip through if the page table entry permits it. With an augmented ISA, software now can express its semantics in a way that hardware understands, e.g., object bounds and types. Hardware can use specialized logic to accelerate costly metadata maintenance and policy check.
Hardbound~\cite{hardbound} 
is based on fat pointers for memory safety. Pointers in the monitored code are extended with a base and bound address pair, allowing hardware to perform bound checks upon each memory access. The compiler and run-time library instrument the monitored code with new instructions to manage bounds. 
Shakti-T~\cite{shakti} 
introduces another level of indirection to reduce the overhead of fat pointer bound loading. Instead of carrying pointer bounds with pointers, they carry indices of pointer bounds, and all pointers having the same index will share a pointer bound. 
HardScope~\cite{hardscope} 
shifts from object bounds to enforcing language variable scoping. A rule stack is maintained by hardware and the compiler instruments the monitored code with new instructions to manage rules. Similar approaches are also applied on COTS computers, e.g., Intel MPX~\cite{mpxexplained}, MPK~\cite{mpk} and ARM Pointer Authentication (which PARTS~\cite{pointerauth} is built on).

\subhead{Information Flow Tracking with tagged memory}
Metadata tags can be associated with memory locations to track dynamic information flow. Tags are propagated by hardware at run-time according to configured policies and anomalies are reacted to also based on policies. Such policies can be specified statically as in Raksha~\cite{raksha} 
and SDMP~\cite{sdmp}, 
e.g., control-flow graph (CFG) passed in at compile time.
Also, new instructions/registers can be introduced to the ISA so that policies can be passed in at run-time with better accuracy and granularity, as in HDFI~\cite{hdfi}. Loki~\cite{loki} is a similar approach but aims to offload access control enforcement to hardware for OS kernels. Security labels of kernel objects are translated into tags by a small trusted monitor and checks are enforced by hardware on each memory access.

\subsubsection{Software participation in metadata management}
While exposing application semantics to hardware is sufficient in most cases, the isolation between the monitored code and hardware-maintained metadata is a double-edged sword. On one hand, it ensures the integrity of metadata even after compromise. On the other hand, hardware must implement all metadata maintenance operations, which can be costly in terms of performance or chip area for hardware. Moreover, certain high-level security policies (e.g. data freshness or confidentiality) may require software support to enforce. 

For such participation, dedicated logic is added to the monitored code ranging from compiler-instrumented instructions to full-fledged run-time libraries exposing developer APIs. In this case, extra care is needed for metadata protection as to how it is isolated from potentially vulnerable code.
Watchdog~\cite{watchdog} 
uses a traditional fat pointer scheme for spatial memory safety, and a lock-and-key approach for temporal memory safety, where each object is associated with a lock, and only pointers with a matching key can dereference the object. Checks and metadata propagation are performed by hardware (there is also the software-only WatchdogLite~\cite{watchdoglite} using compiler instrumentation).
Low-fat Pointers~\cite{lowfat} 
is a fat pointer scheme that uses compressed pointer bound encoding to make it fit into the unused portion of a pointer address. It requires software to place allocated objects properly so that pointer bounds can be efficiently encoded, and encoded pointer bounds are naturally part of pointer values, both in register and in memory. This design avoids the extra memory overhead of pointer bounds, and allows hardware to perform bound check in parallel during pointer dereference.

\section{Discussion} \label{sec:discussion}

\subhead{Relationship between Firmware TEEs}
One observation we have made is that many firmware TEEs are co-dependent on each other, yet are poorly isolated from each other.  While the abstractions between hardware and software create well-defined interfaces, there are no such abstractions between hardware components, a category in which firmware components are also often arbitrarily swept under.  To illustrate this complexity, consider how the Intel ME~\cite{ruan} is blocked from accessing both SMRAM and Intel TXT, as the ME environment has a large TCB and many vectors over which it can be attacked.


Also, the TCBs of most attested TEEs include the firmware TEEs. The original positioning of DRTM (Intel TXT and AMD SVM) was to remove the need for a long chain of trust in SRTM (which included the BIOS for example).  DRTM alleviates this by allowing the start fresh with a minimal TCB after boot (the CPU and optionally a tiny authorized module), which can measure and launch a new VMM/OS---hence the name ``late launch''. 

However, DRTM's TCB does include other firmware (see TCG's DRTM Architecture~\cite{drtm}). For example, SMM is not measured by DRTM for initial integrity, which is why SMI handlers compromised before entering TXT~\cite{attack-txt} could evade DRTM detection and thus take over the whole system (STM~\cite{stm1} is intended to address this). One of Intel SGX's design motivations for running only unprivileged code again was TCB reduction~\cite{sgx}, i.e. to exclude BIOS/UEFI as well as the OS/VMM.  However, even SGX still depends on CPU microcode, another firmware TEE that is a potentially subvertable component.  Furthermore, certain SGX components reside in ME as a trustlet, e.g., the later-added monotonic counter~\cite{ruan}. There have been attempts to add interfaces between firmware components to compartmentalize them.  One example of this is the  SMI Transfer Monitor (STM)~\cite{stm2}, which limits the trust that other components must have in the SMI handlers.  While STM cannot be considered a full abstraction and still requires trust in other firmware TEEs, it at least reduces the attack surface for those firmware TEEs.

\subhead{Towards greater openness}
A significant difference between firmware environments and Attested TEEs is that Attested TEEs are intended for users to deploy custom code, while firmware TEEs are not exposed to users or even software developers (refer to column Dev in Table~\ref{tab:properties}).  This means that the firmware TEEs are largely undocumented, though this naturally does not mean they are any more inherently secure than open TEEs.  This leads to opportunities for research.  

Some of this research leads to changes to improve security, such as SMRR~\cite{ia32-64}. As another example, the group of Koppe and Kollenda have used reverse engineered microcode to change the behavior of earlier CPU models (specifically AMD K8 and K10)~\cite{microcode2,microcode3}.  An interesting direction for this line of work is the standardization of firmware access, which could lead to greater customization and modification of hardware~\cite{microcode3}.

\section{Concluding Remarks}
We explain the perceived advantages/disadvantages of hardware security features, and the root cause of the notorious low-level attacks (firmware or side-channel), using the  concept of abstraction, which is ubiquitous in computing. We find that abstraction either underdone or overdone can lead to 
vulnerabilities, many of which have commonalities across various TEEs.  We draw a lesson that future researchers and secure hardware designers may do well to examine their abstractions, not just for ease of programmability, maintenance and performance, but for whether they are fulfilling their security requirements in terms of the information flows they allow.

\balance


\vfill\null
\appendix
To facilitate understanding of Table~\ref{tab:armx86z}, a brief introduction to the acronyms and terms used is provided as follows.

\subsection{Glossary}

\emph{Processor Units}: 

\textbf{SAP}: System Assist Processor.

\textbf{zIIP}: Integrated Information Processor.

\textbf{IFP}: Integrated Firmware Processor.

\textbf{IFL}: Integrated Facility for Linux.

\textbf{ICF}: Internal Coupling Facility.

\vspace{5pt}
\emph{Stand-alone components}: 

\textbf{CU}: Control Unit. A CU consolidates device-specific logic and circuitry and acts as a hub between the main system and multiple devices of the same type.

\textbf{CSS}: Channel Subsystem. I/O operations are offloaded to the CSS (saving the main processors from waiting), allowing the handling of massive numbers of concurrent transactions. The CSS runs on a dedicated processor SAP (System Assist Processor). It is comparable to the SouthBridge on an x86 motherboard but more full-fledged and stand-alone.

\textbf{SE}: Support Element. SE has the lowest-level access to mainframe processors via the processor interconnect network. It is a console as the single-point-of-control to the mainframe. The SE takes the physical form of a Thinkpad laptop.

\textbf{HMC}: Hardware Management Console is a hardware appliance with which various configuration and administrative operations can be performed. It can be emulated with a Web application on a PC.

\subsection{Additional Information on TEEs of Different Platforms} 

\emph{sFW on ARM}. Due to the diversity of ARM platforms, the implementation of individual SoCs (System-on-Chips) is vendor-specific. This is exemplified by bootloaders, e.g., Samsung S3 had PBL loading SBL1, SBL2 and SBL3, followed by ABOOT~\cite{android}.

\emph{sFW on mainframes}. As an example among many other firmware TEEs, PR/SM (Processor Resource/System Manager) is a type-1 hypervisor implemented as firmware capable of directly partitioning physical processors, memory and I/O channels.

\emph{Applications on co-processors}.
Intel ME has become a platform for deployment of security-critical applications therein (trustlets). The same applies to AMD PSP (Platform Security Processor), or AMD Secure Processor, as well, omitted here for brevity. Examples include Active Manageability Technology (AMT) for out-of-band remote access, Protected Audio/Video Path (PAVP) and firmware-based TPM (fTPM). 

Examples are not provided for the baseband processor, Apple Secure Enclave Processor, System Control Processor (SCP) and Manageability Control Processor (MCP), for the same diversity reason in sFW.


\end{document}